\documentclass[11pt]{article}
\usepackage{amsmath,bm,bbm}
\usepackage{geometry,epsf}
\usepackage{hyperref}
\newcommand{\BM}[1]{{\mbox{\boldmath{$#1$}}}}

\def\gtwid{\mathrel{\raise.3ex\hbox{$>$\kern-.75em\lower1ex\hbox{$\sim
$}}}}
\def\vio{\mathrel{\hbox{$R$\kern-.60em\hbox{$/
$}}}}
\usepackage{graphicx}
\usepackage{amssymb}
\def\lsim{\mathrel{\raise.3ex\hbox{$<$\kern-.75em\lower1ex\hbox{$\sim$}}}}
\def\gsim{\mathrel{\raise.3ex\hbox{$>$\kern-.75em\lower1ex\hbox{$\sim$}}}}

\begin{document}

\title
{Electromagnetic properties of dark matter:\\ dipole moments and charge form factor}
\author{Vernon Barger$^{1}$, Wai-Yee Keung$^{2}$, Danny Marfatia$^{3}$\\[2ex]
\small\it ${}^{1}$Department of Physics, University of Wisconsin, Madison, WI 53706, U.S.A.\\
\small\it ${}^{2}$Department of Physics, University of Illinois, Chicago, IL 60607, U.S.A.\\
\small\it ${}^{3}$Department of Physics and Astronomy, University of
Kansas, Lawrence, KS 66045, U.S.A.}

\date{}

\maketitle

\begin{abstract}
A neutral dark matter particle may possess an electric
dipole moment (EDM) or a magnetic dipole moment (MDM), so that its
scattering with nuclei is governed by electromagnetic interactions.
If the moments are associated with relevant operators of dimension-5, they may 
be detectable in direct search experiments. We calculate
complete expressions of the scattering cross sections and the recoil energy
spectra for dark matter with these attributes.
We also provide useful formulae pertinent to dark matter that
interacts via an electric charge form factor (CFF) which is related to the charge radius
defined by an effective dimension-6 operator. We show that a 7~GeV dark matter particle 
with an EDM, MDM or CFF easily reproduces the CoGeNT excess while remaining consistent with
null searches.

\end{abstract}
\newpage
\section{Introduction}

The nature of the dark matter (DM) particle is unknown.  There
are many well studied scenarios guided by theoretical niceties such
as the Minimal Supersymmetric Standard Model, models with extra dimensions, and the 
Little Higgs Models with $T$-parity, all of which posit that the DM particle
interacts primarily via weak interactions. On the experimental side, 
direct DM searches are looking for signals of the
recoiling nuclei from DM-nucleus scattering. We
take the uncommon view that the scattering process may
be electromagnetic in nature. The interaction occurs through the
electric dipole moment (EDM) or the magnetic dipole moment (MDM) of
dark matter. The interactions are described by dimension-5 operators for 
non-self-conjugate particles, such as Dirac DM, but not for Majorana
DM. The EDM and MDM of DM can be induced by underlying short distance
physics at the one-loop order or higher.  As there are no strong
reasons against large CP violation in the DM sector, we cannot neglect
the EDM possibility. In fact, for comparable short distance cutoffs,
we find that an EDM may give the dominant contribution to DM-nucleus
scattering because it directly couples to the nuclear charge.  
In the EDM case, the recoil energy $E_R$ distribution is highly
enhanced in the low $E_R$ region as $1/(v_r^2 E_R)$ (where $v_r$ is the
speed of the DM particle in the rest frame of the nucleus) 
in contrast with that from MDM which goes as $1/E_R$.

While DM-nucleus scattering due to the MDM of DM has been studied extensively
in the literature, the relevant formulae
with the correct dependence on the nuclear charge $Z$ and nuclear
moment $\mu_{Z,A}$ are not available. In this work,
we provide analytic expressions for
 the scattering cross sections for the MDM and EDM cases with careful
expansions in the relative velocity and recoil energy.
To complete the treatment of electromagnetic properties of DM, we extend our
analysis to the dimension-6 operator, which is the electric charge form factor (CFF)
slope or the charge radius of the neutral DM. The operator has a structure that is
similar to that of the spin-independent (SI) interaction.


\section{Electric dipole moment of dark matter}
The effective non-relativistic Hamiltonian of the EDM of a particle with spin $\BM{S}$ 
is 
$$    H_{\rm eff}=-{\tt d} \BM{E}\cdot\BM{S}/S  \,, $$
with the normalization chosen to agree with the standard form for a spin ${1\over2}$
particle, {\it i.e.}, $-{\tt d} \BM{E}\cdot\BM{\sigma}$, where ${\tt d}$ is dimensionful, and the electromagnetic energy density is $(E^2+B^2)/2$.
As the electric field of the nucleus $\BM{E} =  -\hbox{ grad }\phi$, 
we identify the gradient with the momentum transfer $\BM{q}$.
Therefore, the  direct scattering of the DM particle 
$\chi$ and the nucleus $N$,
$\chi N \to \chi N$, 
via the interaction between the DM  electric dipole (with moment ${\tt d}_\chi$) 
and the nuclear charge $Ze$ (with $e^2=4\pi\alpha$) is
$$
{\cal M}=
\BM{S} \cdot i\BM{q} \ 
\left({ {\tt d}_\chi\over S} {1\over \BM{q}^2} Ze \right)\,. $$
It is important to note that the momentum transfer $\BM q$ is Galilean
invariant and is conveniently related to the center-of-mass momentum $\BM p$.
Hence
$ d\BM{q}^2=2 |\BM{p}|^2 d\cos\theta  =4 |\BM{p}|^2 {d\Omega\over4\pi}$,
and $|\BM{p}|=m_r v_r$, where
$m_r \equiv {m_A m_\chi \over m_A+m_\chi}$ is the reduced mass; 
$m_A$ and $m_\chi$ are the masses of the nucleus and
DM particle, respectively.
$\BM{q}^2$ ranges from 0 to $(2m_rv_r)^2$, as is easily checked 
in the center-of-mass frame in which the momenta have equal magnitudes $m_r v_r$.\footnote{The DM particle velocity in the frame of the galactic halo
is usually described~\cite{Jungman:1995df}  by a Maxwellian distribution,
$$    f^G(\BM v) d^3\BM{v} 
={N\over (v_0 \sqrt{\pi})^3} e^{-v^2/v_0^2} d^3\BM{v}\,, $$
where the most probable velocity is $v_0=230$~km/s, and the
distribution is cut off at the escape velocity \mbox{$v_{\rm esc}=600$~km/s}. 
The normalization $N$ is close to 1, or more precisely,
$$ N= {1\over {\rm erf}(v_{\rm esc}/v_0)
              -{2\over\sqrt{\pi}}(v_{\rm esc}/v_0)\exp(-{v_{\rm esc}^2/v_0^2)}   
             } \,.$$
The one-variable velocity distribution is
$$  f^G_1(v) dv ={4v^2N\over v_0^3\sqrt{\pi}}e^{-v^2/v_0^2} dv\,. $$ 
However, since the solar system is moving at a speed $v_E=244$ km/s
with respect to the halo~\cite{speedref} and  $v_E \sim v_0$, we need to
use a  more relevant  distribution of the relative velocity,
$\BM{v}_r=\BM{v}-\BM{v}_E$.
As an approximation we ignore the seasonal motion
of the Earth around the Sun with a relative speed of about 30~km/s. 
Then,
$$ f(\BM{v}_r) d^3\BM{v}_r
=f^G(\BM{v}) d^3\BM{v}   
=f^G(\BM{v}_r+\BM{v}_E) d^3\BM{v}_r   \quad
\hbox{because }
d^3\BM{v}=d^3\BM{v}_r   \ .$$
For a fixed $v_r$, we integrate the polar angle between $\BM v_r$ and $\BM v_E$ to obtain
$$ f_1(v_r) dv_r=N
{v_r dv_r\over v_E v_0 \sqrt{\pi}}
\left(e^{-(\min(v_r-v_E,v_{\rm esc}))^2/v_0^2}
-e^{-(\min(v_r+v_E,v_{\rm esc}))^2/v_0^2} \right) \,. $$
If $v_r> v_{\rm esc}+v_E$, $f_1(v_r)=0$;  
see Refs.~\cite{Jungman:1995df, Drees:2007hr}.}
We use the spin relation,
$$ {\rm Tr}( S^i S^j)  = (2S+1) \delta^{ij} S(S+1)/3  \ ,$$
to obtain the spin-averaged differential cross section,
\begin{equation}
 d\sigma_{EDM}(\chi N) = {1\over 4\pi}
 {\tt d}_\chi^2 Z^2 e^2  {(S+1) \over 3 S}
{1\over  v_r^2} {d\BM{q}^2\over \BM{q}^2} |G_E(\BM{q}^2)|^2\,.
\label{sigEDM}
\end{equation}
We have included the nuclear charge form factor
$|G_E(\BM{q}^2)|^2$ 
to incorporate elastic scattering effects off a heavy nucleus.\footnote{A good nuclear form factor 
can be found in Ref.~\cite{NucForm}. The spatial charge distribution is parameterized by the Fermi distribution $ \rho({\bf r})=\rho_0/(1+e^{(r-c)/a_0}) $, where  the radius at which the density is $\rho_0/2$ 
is  $c=(1.18 A^{1\over3}-0.48)$ fm and the edge
thickness parameter is $a_0=0.57$~fm~\cite{preston}.
The form factor that is valid for nuclei with a well-developed core ({\it i.e.}, with atomic masses above 20) is obtained by the Fourier transform in the limit
$c\gg a_0$, 
$$  G_E(q)=\left[{\pi a_0\over c}{\sin(qc)\cosh(\pi a_0 q)
                                \over  \sinh^2(\pi a_0 q)}
                                -{\cos(qc)\over \sinh(\pi a_0 q)}
         \right]{4\pi^2\rho a_0c \over q} \,, \quad \ \ \rho_0={3\over 4\pi c^3}{1\over 1+(a_0\pi/c)^2}\,.  $$
Note that $G_E(0)=1$.}
Accounting for a difference in convention in the definition of $e$, 
our formula agrees with that of Ref.~\cite{Pospelov:2000bq}.

We relate $\BM{q^2}$ to the nuclear recoil energy in
the lab frame, $\BM{q^2}=2m_A E_R$, and find
\begin{equation}
 {d\sigma_{EDM}\over dE_R}= {1\over 4\pi}
 {\tt d}_\chi^2 Z^2 e^2  {(S+1) \over 3 S}
{1\over  v_r^2} {1\over E_R} |G_E(\BM{q}^2)|^2\,.
\end{equation}
The $1/(v_r^2 E_R)$ dependence is characteristic of the EDM of the DM particle.\footnote{The differential reaction rate (per unit detector mass) is
$$ {dR\over dE_R}
={\rho_0\over m_\chi}{1\over m_A} 
\int_{v_{\rm min}}^\infty 
v_r f_1(v_r) {d\sigma\over dE_R} dv_r \,,$$
where the local DM density $\rho_0=0.3$ GeV/cm$^3$ and $v_{\rm min}=\sqrt{m_A E_R \over 2 m_r^2}$.
$dR/dE_R$ includes contributions from both $\chi$ and its conjugate $\bar\chi$
for they have the same cross sections.

In the non-relativistic limit, the differential cross section can be
Maclaurin expanded in powers of $v_r$. The two most important
contributions are
$$ d\sigma \sim {1\over v_r^2} d \{ \sigma_- \}
+ d \{  \sigma_+ \} \ ,$$ 
with $v_r$ independent coefficients denoted by  brackets. (For example, in Eq.~\ref{sigEDM},
$d\{\sigma_-\}$ is the coefficient of $v_r^{-2}$, and $d\{\sigma_+\}=0$.) 
Usually, the first term is the relevant one (as in
the EDM, CFF, or SI cases). However, in certain cases
like MDM, the second term may compete due to the
$1/E_R$ enhancement from  the low energy virtual photon
propagator. On integrating, we find
$$ {dR\over dE_R}
={\rho_0\over m_\chi}{1\over m_A} 
 \left[  {d \{ \sigma_- \}\over dE_R} {1\over v_0} I_-
+        {d \{\sigma_+\}\over dE_R}  v_0 I_+ \right]\,, $$
where the dimensioless integrals are defined by
$$ {I_-\over N}={v_0\over 2v_E}\left[
 {\rm erf}\left({v_u\over v_0}   \right)
-{\rm erf}\left({v_d\over v_0}   \right)
-{2\over\sqrt{\pi}}\left({v_u\over v_0} -{v_d\over
 v_0}\right)e^{-v_{\rm esc}^2/v_0^2}
               \right] \,,$$
and 
$$ {I_+\over N}= 
\left({v_d\over 2v_E\sqrt{\pi}}+{1\over\sqrt{\pi}} \right) 
 e^{-v_d^2/v_0^2}-
\left({v_u\over 2v_E\sqrt{\pi}}-{1\over\sqrt{\pi}} \right) 
e^{-v_u^2/v_0^2}$$
$$+{v_0\over 4 v_E} \left({1+{2v_E^2\over v_0^2}}\right) 
\left({\rm erf}\left({v_u\over  v_0}  \right) 
     -{\rm erf}\left({v_d\over  v_0}  \right)\right) $$
$$
\quad -{1\over\sqrt{\pi}}\left[2+
{1\over 3v_E v_0^2}\left(
    (v_{\rm min}+v_{\rm esc}-v_d)^3
  - (v_{\rm min}+v_{\rm esc}-v_u)^3
                   \right)
                    \right]e^{-v_{\rm esc}^2/v_0^2}\,,
$$
with the shorthand $v_u=\min(v_{\rm min}+ v_E,v_{\rm esc})$,
$v_d=\min(v_{\rm min}- v_E,v_{\rm esc})$.
Note that $I_-=0$ for $v_{\rm min}> v_{\rm esc}+v_E$.} 

To have an EDM, the DM particle cannot be self-conjugate. Consequently, for
$S={1\over2}$, the particle has to be Dirac.
Note that the spin factor $S+1\over 3S$ becomes 1 for $S={1\over2}$
in our  numerical illustrations. 
Our result also applies to the anti-dark matter particle under the assumption that CPT is conserved.

\section{Magnetic dipole moment of dark matter}
In the static limit, the DM magnetic moment can only couple to the
nuclear magnetic moment via the induced magnetic field.
The  non-relativistic effective Hamiltonian of a magnetic 
moment  of a particle with spin $\BM{S}$ subject to a magnetic field $\BM{B}$ 
is, in our convention
$$    H_{\rm eff}=-{{ \mu }} \BM{B}\cdot\BM{S}/S 
 \,,$$
which agrees with the standard form for a spin ${1\over2}$
particle, {\it i.e.}, $-{{ \mu }}\BM{B}\cdot\BM{\sigma}$, where $\mu$ is dimensionful.
Since  $\BM{B} =  \BM\nabla \times\BM{A}$, 
we identify the curl $\BM\nabla \times$ 
with the momentum transfer $\BM{q} \times$~.
The  direct scattering off the nucleus of spin $I$  via the interaction between the
nuclear magnetic moments ${{ \mu }_{Z,A}}$ and  the magnetic 
moment ${{ \mu }_\chi}$ of the DM particle is described by 
$$ 
{\cal M}=
\BM{S} \times \BM{q} \cdot 
\left({ {{ \mu }_\chi}\over S} {1\over \BM{q}^2} {{{ \mu }_{Z,A}}\over I}\right)
\BM{I}\times \BM{q}   \ .$$
The curl substitution has been  applied to the DM vertex and the nuclear
vertex which are linked by the virtual photon exchange
factor $1/\BM{q}^2$. 
The spin-averaged differental cross section is
\begin{equation}
 d\sigma_{MDM}(\chi N) = {2\over \pi}
 {{ \mu }^2_\chi} {{ \mu }^2_{Z,A}} 
{S+1 \over 3S} 
{I+1 \over 3I} 
{d\BM{q}^2\over 4v_r^2}   |G_M(\BM{q}^2)|^2\,.
\end{equation}
The magnetic nuclear form factor $G_M$ has been included. 
The above purely magnetic description ignores an important nuclear charge
$Z^2$ effect which is suppressed by $v_r^2$ but enhanced by $1/E_R$.
Both the magnetic and electric effects are comparable 
at direct search experiments with low recoil energy thresholds.

To account for the $Z^2$ effect, we need to treat the convection current of the
nucleus \mbox{$Z(p_{Z,A}^\mu+{p'}_{Z,A}^\mu)$} (where $p_{Z,A}^\mu$ and ${p'}_{Z,A}^\mu$
are the incoming and outgoing momenta, respectively) with a Dirac trace calculation.
Then we include the above contribution
from the nuclear magnetic moment associated with its spin $I$.  
The leading contributions for low 
recoil energy and low relative velocity give
\begin{equation} 
{d\sigma_{MDM}\over d E_R}
={e^2 \mu^2_\chi\over 4\pi E_R}{S+1\over 3S}
\left[ Z^2 \left(1-{ E_R\over 2 m_A  v_r^2}-{ E_R\over  m_\chi  v_r^2}
\right) |G_E|^2 
+ {I+1\over 3I}\left( {\mu_{Z,A}\over {e\over 2m_p}} \right)^2 
{m_A E_R\over  m_p^2 v_r^2} |G_M|^2
\right] \,.
\end{equation}

The two nuclear form factors $G_E$ and $G_M$ are normalized so that
$G_E(0)=1$ and $G_M(0)=1$; the charge  $Z$ and the moment
$\mu_{Z,A}$ have been factored out.
In our numerical analysis we make the simplifying assumption
that the two nuclear form factors $G_E$ and $G_M$
are approximately equal, although they could be slightly different.

Several studies only include the nuclear spin and ignore the effect of the nuclear 
charge, {\it e.g.}, Ref.~\cite{Pospelov:2000bq}.
Others inappropriately scale the result for a Dirac point charge by 
$Z^2$~\cite{Sigurdson:2004zp,Masso:2009mu}. 
Our result reduces to that of Ref.~\cite{Masso:2009mu} in the limit that a nucleus becomes a Dirac point charge, $Z\to 1$,
${\mu_{Z,A}/ {e\over 2m_p}} \to 1$, $m_A\to m_p$, and $I\to 1/2$, and has subsequently been reproduced in an erratum to Ref.~\cite{Cho:2010br}.
It is worth pointing out that a proton with a significant anomalous magnetic moment is not faithfully represented by a Dirac point charge. Note that the $A^2$ factor in the result of 
Ref.~\cite{bagnasco} must be replaced by $(m_A/m_p)^2$
to agree with our expression, and for quantitative accuracy.

\section{Electric charge form factor of dark matter}
A neutral non-self-conjugate dark matter particle  can
couple directly to the photon via its charge form factor (CFF)
$G_\chi(\BM{q}^2)$; charge neutrality only implies $G_\chi(0)=0$. 
At low momentum transfer, it is conveniently parameterized by a cutoff
$\Lambda_{CFF}$:
$$ G_\chi(\BM{q}^2) =\BM{q}^2/\Lambda_{CFF}^2  \ . $$
The amplitude for DM-nucleus scattering via the form factor is given by 
$$ {\cal M}= G_\chi(\BM{q}^2) (e^2/\BM{q}^2) Z G_E(\BM{q}^2) \ , $$
which is clearly spin-independent,
so that the differental cross section is
\begin{equation}
 d\sigma_{CFF}(\chi N) = {e^4 |ZG_E(\BM{q}^2)|^2 \over \pi\Lambda_{CFF}^4}
{d\BM{q}^2\over 4v_r^2} \,. 
\end{equation}
Employing the cutoff $\Lambda_{CFF}$ 
translates into a useful cross section for DM-proton scattering:
$$ \sigma^{(p)}_{CFF}={1\over\pi} {m_p^2 m_\chi^2\over (m_\chi+m_p)^2} 
\left({e^2\over\Lambda_{CFF}^2}\right)^2
\simeq 
\left({1 \hbox{ TeV }\over \Lambda_{CFF}}\right)^4
{ m_\chi^2\over (m_p+m_\chi)^2} \times 0.92  \times 10^{-42}\,
  \hbox{cm$^2$ ,}
$$
which is relevant to the DM capture rate in the Sun. The proton form
factor is essentially flat and ignored during the integration.

In comparison, the SI scattering amplitude,
$$ {\cal M}_{SI}
= 2 [f_pZ+f_n(A-Z)] G_E(\BM{q}^2) \,,$$
(where we have used the electric form factor to approximate the
nuclear size effect), 
yields the differential  DM-nucleus cross section,
$$ {d\sigma^{(Z,A)}_{SI}\over dE_R}
= {4\over\pi} [f_pZ+f_n(A-Z)]^2 |G_E(\BM{q}^2)|^2 {m_A\over 2v_r^2}\,,$$
and the DM-proton SI cross section, 
$$ \sigma^{(p)}_{SI}=\left({Z\over A}\right)^2 \sigma^{(p)}_{CFF}={4\over\pi} {m_p^2 m_\chi^2\over (m_\chi+m_p)^2} f_p^2\,.$$
We therefore have the correspondence,
$$  {e^2\over\Lambda_{CFF}^2}Z  \longleftrightarrow     2 [f_pZ+f_n(A-Z)]\,.  $$
If we assume $f_p=f_n=\Lambda^{-2}_{SI}$ for simplicity, we can further
simply to the correspondence,
$$ {1\over \Lambda^2_{SI}}  
\longleftrightarrow  
{1\over2 }\left({Z\over A}\right) {e^2\over  \Lambda^2_{CFF}} \,.$$
Since $2Z\approx A$ for most nuclei, 
the correspondence becomes
$$ 
\Lambda_{SI}\leftrightarrow {2\over e}\Lambda_{CFF}\approx 6.6\ \Lambda_{CFF}\,.
$$

\section{Analysis}

For the MDM of DM, we define the dimensionless gyromagnetic ratio or g-factor ${\tt g}_\chi$ by 
$${\tt g}_\chi S ={{ \mu }_\chi \over {e\over 2 m_\chi}}\,.$$
It is useful to compare ${\tt d}_\chi$ and 
${ \mu }_\chi$
in terms of the effective cutoff scales  $\Lambda_{EDM}$ and $\Lambda_{MDM}$  
defined by
$$ {\tt d}_\chi \equiv {e \over\Lambda_{EDM}}   \quad  \hbox{ and } \quad  
  { \mu }_\chi={\tt g}_\chi S {e\over 2 m_\chi} 
       \equiv {e\over\Lambda_{MDM}}  \ . $$
Then,
$$ {\Lambda_{EDM} \over  1\hbox{ TeV} } 
= 19733\times { 10^{-21}\  e\cdot\hbox{cm} \over {\tt d}_\chi } \quad
\hbox{ and} \quad
{\Lambda_{MDM}\over  1\hbox{ TeV}}  = {2\over {\tt g}_\chi S } 
{m_\chi \over 1\,\hbox{TeV}} 
\ .
$$ 
(Note that the cutoff scales are arbitrarily defined quantities, related to compositeness
or short distance physics. They are defined in order to facilitate comparison and are not to
be interpreted as new physics scales.)
Direct DM search experiments are currently sensitive to ${\tt d}_\chi \sim 10^{-21}$ $e\cdot$cm which translates
into a sensitivity to $\Lambda_{EDM}$ of almost 20~PeV. On the other
hand, experiments are sensitive to ${\tt g}_\chi S\sim 0.001$ or $\Lambda_{MDM}$ of order 10 TeV
for $m_\chi \lsim 10$~GeV.
The difference arises because the MDM contribution to
direct searches is more non-relativistically  suppressed.
Due to the higher dimensionality of the DM charge form factor, 
the corresponding scale $\Lambda_{CFF}$ can be probed to about a few 100~GeV.

We now consider the consequences of these electromagnetic properties of DM for
direct searches.\footnote{The relevant nuclear parameters for the CoGeNT, CDMS and XENON
experiments are~\cite{nuclparm},
\begin{center}
$_{32}^{73}$Ge:  $I={9\over2}$, 
    ${ \mu_{Z,A} }=-0.8795 {e\over 2  m_p}$,
    $ f=7.73\%$\\ 
\noindent
$_{54}^{129}$Xe: $I={1\over2}$, 
    ${ \mu_{Z,A} }=-0.778 {e\over 2  m_p}$, $f=26.44\%$\\
$_{54}^{131}$Xe: $I={3\over2}$, ${ \mu_{Z,A} }=0.692 {e\over 2 m_p}$,
    $f=21.18\%$\,,
\end{center}
where $f$ is the natural abundance of the isotope.
}
Since the DM velocity cannot exceed $v_{\rm esc}$, which we take to be 600~km/s, 
a DM particle lighter than 7.2~GeV 
cannot give a nuclear recoil energy above~10 keV, which was the
threshold cut  adopted by the CDMS II collaboration~\cite{Ahmed:2009zw}. 
Although CDMS II  has accumulated 
more kg$\cdot$days of data than CoGeNT~\cite{Aalseth:2010vx}, the lower threshold energy of
CoGeNT permits sensitivity to masses as low as 3~GeV. Solely for the purpose of
illustration, we focus on the CoGeNT anomaly.

\begin{figure}[t]
\label{cog} 
\begin{center}
\leavevmode
\epsfxsize=5in
\epsfbox{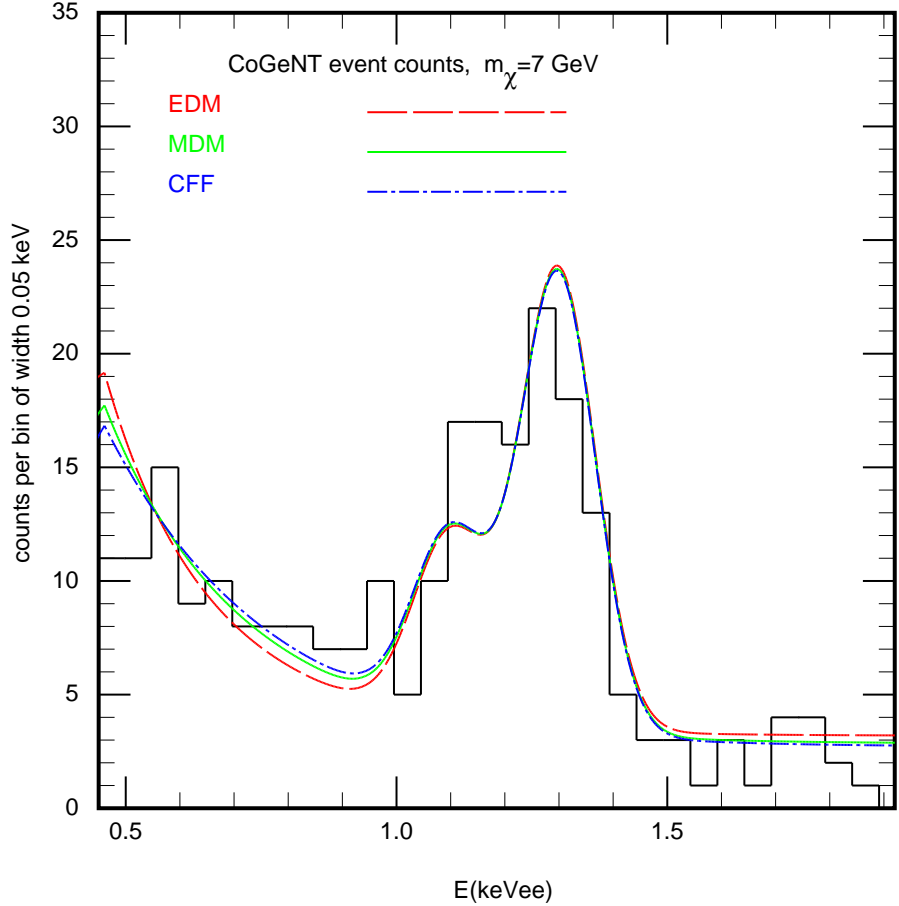}
\caption{
A DM particle with $m_\chi=7$~GeV and an EDM, MDM or CFF reproduces CoGeNT data from a 
56-day run with 0.33~kg of germanium.
Note that the shape of the CFF curve is identical to that for spin-independent scattering.
}
 \end{center}
\end{figure}

In Fig.~1, we show that a 7~GeV DM particle with an EDM of ${\tt d}_\chi=10^{-20}$ $e\cdot$cm (equivalently 
$\Lambda_{EDM}=1.97$~PeV), or an MDM with 
${\tt g}_\chi S=0.00454$ (equivalently $\Lambda_{MDM}=3.09$~TeV), or a CFF with $\Lambda_{CFF}=187$~GeV
(corresponding to $\sigma^{(p)}_{CFF}=5.8\times 10^{-40}$~cm$^2$ or $\sigma^{(p)}_{SI}=1.1\times 10^{-40}$~cm$^2$) 
easily reproduces the CoGeNT event excess below 2~keVee; we employed
an energy-dependent quenching factor, $E({\rm{keVee}})=0.19935\,E_R(\rm{keV})^{1.1204}$,
which converts the total nuclear recoil energy $E_R$ to the energy detected by the experiment (in the form of ionization, scintillation or heat) with units 
of equivalent electron energy (keVee).
The corresponding $\chi^2$ values for the 3-parameter fit to 30 data points between 0.45~keVee and 1.9~keVee 
are $\chi^2_{EDM}=26$, $\chi^2_{MDM}=22$ and $\chi^2_{CFF}=20$, where~\cite{pdg}
$$\chi^2= \sum_{i\ \rm{with}\ N_i^{exp} \neq 0} 2(N_i^{th}-N_i^{exp}+N_i^{exp} \ln {N_i^{exp}\over N_i^{th}}) + \sum_{i\ \rm{with}\ N_i^{exp}=0} 2 N_i^{th}\,,$$
and the $N_i^{th}$ include a background contribution that is modeled as a linear combination of a constant term
and a sum of two weighted Gaussian distributions that describe the peaks from the decay of $^{65}$Zn and $^{68}$Ge
via L-shell electron capture~\cite{Aalseth:2010vx}. Two of the fit parameters fix the background and the third parameter normalizes the
signal. In all cases, the constant background is about 3 events/bin which explains the the data between 
1.5~keVee and 3.2~keVee satisfactorily.

Since in all cases the DM-nucleus scattering is dominantly SI , the most stringent constraints in this low $m_\chi$ range are obtained from the XENON10 experiment 
due its low energy threshold~\cite{xe10} (and perhaps CDMS-II silicon detector data~\cite{sili} which may have
an underestimated energy calibration uncertainty). 
We assert that the DM candidates of Fig.~1 are 
consistent with the XENON10 data based on the following: (1) The shape of the CFF recoil energy 
distribution is identical to that for SI scattering, and depending on analysis details, 
SI scattering of a 7~GeV DM particle with $\sigma^{(p)}_{SI} \sim 10^{-40}$~cm$^2$ is either compatible with~\cite{hooper} or 
marginally excluded by~\cite{sorensen} XENON10 data, (2) the distributions 
of Fig.~1 are almost identical, and (3) we checked that for $m_\chi=7$~GeV, the difference in the recoil energy distributions
for CFF scattering on Ge and Xe is larger than for scattering via an EDM or MDM.  
Taking these facts together, we infer that the DM candidates of Fig.~1 are not in conflict with XENON10 data.

\section*{Acknowledgments}

We thank J.~Collar for providing us with a wee note on CoGeNT data. This work was supported by DoE Grant Nos. DE-FG02-84ER40173, 
DE-FG02-95ER40896 and DE-FG02-04ER41308, by NSF
Grant No. PHY-0544278, and by the Wisconsin Alumni Research Foundation. 
D.M. thanks the Aspen Center for Physics for its hospitality.
\appendix

\end{document}